\documentclass[aps,prb,amssymb,showpacs,twocolumn,longbibliography,superscriptaddress,nodois]{revtex4-2}

\usepackage{graphics,graphicx,amsmath}
\usepackage{tabularx,float}
\usepackage{epsfig}
\usepackage{subfigure}
\usepackage{bm}
\usepackage{stmaryrd}
\usepackage{mathtools}
\usepackage{wasysym}
\usepackage{bbm}
\usepackage[T1]{fontenc}
\usepackage[french,english]{babel}

\usepackage{mathrsfs} 
\usepackage{physics}

\usepackage{color}
\usepackage{verbatim}
\graphicspath{{fig/}}

\definecolor{myblue}{RGB}{45, 48, 146}
\usepackage[colorlinks=true,
            linkcolor=myblue,
            citecolor=myblue,
            urlcolor=myblue]{hyperref}

\newcommand{\pd}[1]{\partial_{#1}}
\newcommand{\mc}[1]{\mathcal{#1}}
\newcommand{\Ket}[1]{\vert#1\rangle}

\newcommand{\Braket}[2]{\langle#1\vert#2\rangle}

\newcommand{\mf}[1]{\mathbf{#1}}
\newcommand{\bk}{\mathbf{k}}
\newcommand{\bd}{\mathbf{d}}
\newcommand{\pmu}{\partial_\mu}
\newcommand{\pnu}{\partial_\nu}

\DeclareMathOperator{\sgn}{sgn}

\begin{document}

\date{\today}

\title{Composite quantum geometry of superconductors}

\author{Florian Simon}
\affiliation{Department of Physics and Astronomy, Uppsala University, Box 524, 75120 Uppsala, Sweden}
\affiliation{Technical University of Munich, TUM School of Natural Sciences, Physics Department, 85748 Garching, Germany}
\affiliation{Munich Center for Quantum Science and Technology (MCQST), Schellingstr.~4, 80799 M{\"u}nchen, Germany}
\author{Thomas Bernat}
\affiliation{Department of Physics and Astronomy, Uppsala University, Box 524, 75120 Uppsala, Sweden}
\author{Annica M. Black-Schaffer}
\affiliation{Department of Physics and Astronomy, Uppsala University, Box 524, 75120 Uppsala, Sweden}

\begin{abstract}
    The interplay of superconductivity and the quantum geometry of the normal state has recently been the subject of an array of studies, especially regarding the superfluid weight. In this work, we turn our attention to the quantum geometry of the superconducting state itself, set by the Bogoliubov-de Gennes (BdG) Hamiltonian, which dictates the geometric and topological properties of superconductivity. We show that under three general conditions, namely superconducting fitness, orbital uniformity of the superconducting pairing, and absence of normal-state spin-flip terms, the BdG quantum geometry exactly separates into a sum of the normal-state quantum geometry and an additional pairing quantum geometry, thereby displaying a simple composite structure. We show that this separation holds for all spin-singlet and -triplet pairings, including nonunitary spin-triplet pairing. We further provide explicit analytical formulas for the pairing quantum geometry for all these cases. These results establish how superconducting pairing alone easily drives both topology and a finite quantum metric, thus being present even in topological trivial or flat band superconductors, with no normal state quantum geometry.
    To complement these results, we also derive the BdG quantum geometry of a general two-orbital spin-singlet superconductor with non-uniform pairing and finite superconducting fitness. Here, our explicit analytical results establish a non-separable composite BdG quantum geometry, with the normal state and pairing contributions generally intertwining, thereby producing even more possibilities for finite quantum geometry. 
    Our results provide design rules for creating superconductors and superconducting hybrid structures with nontrivial topology and finite quantum metric and will additionally help in the experimental diagnosis of unconventional superconductivity. 
\end{abstract}

\maketitle
\section{Introduction}
Quantum geometry has recently come to play a key role in understanding advanced materials and phases of matter \cite{Cayssol2021,Liu2025,Yu2025}. 
It characterizes the geometric structure of parameter-dependent quantum states through the quantum geometric tensor (QGT), and becomes important as soon as there are more than one low-energy orbital present, as is often the case in real materials. Since the real and imaginary parts of the QGT correspond to the quantum metric and Berry curvature, respectively \cite{Provost1980,Berry1989,Cayssol2021}, it encodes both the distance in quantum state space and the topology. In crystalline solids, the momentum dependence of Bloch eigenstates endows electronic bands with a quantum geometry \cite{Simon2013, Cayssol2021}, which underlies a wide range of phenomena. The Berry curvature is the core part of topological band theory and governs responses such as Hall transport and electric polarization \cite{Hasan2010,Ando2013,Xiao2006,Oka2009,Xiao2010,Liu2025,Jiang2025,Gao2025}, as well as produces boundary states, while the quantum metric of the bands has more recently emerged as an essential contribution to diverse physical observables \cite{Komissarov2024,Yu2024,Liu2025,Jiang2025,Sala2026}. 

Conceptually similar band theory applies to superconductors (SCs), effectively described within the Bogoliubov–de Gennes (BdG) formalism, where bands of quasiparticle excitations of the Cooper pair condensate arise from the hybridization of electron and hole degrees of freedom \cite{Bogoljubov1958,Tinkham2004,Sato2017}. Their topology has already been the subject to intense research as it offers tantalizing prospects of robust quantum computers \cite{Kitaev2001,Alicea2012,Sato2017,Mandal2023}. In contrast, their full quantum geometry, hereafter referred to as \textit{BdG quantum geometry}, has received comparatively little attention \cite{Porlles2023,Keskiner2026}. Existing works have instead largely focused on how the quantum geometry of the normal state influences certain SC properties \cite{Peotta2015,Iskin2018,Kitamura2022,Iskin2023,Yu2024,Daido2024,Hu2025,Simon2025,Kitamura2026}, especially the superfluid weight in flat band SCs. This is despite it being the BdG quantum geometry, and not the quantum geometry of the normal state, that determines the intrinsic structure and properties of the SC state.

Notably, although BdG Hamiltonians can be treated as effective Bloch Hamiltonians, the two are fundamentally different. Bloch Hamiltonians describe the normal state, associated with elementary electrons, whereas BdG Hamiltonians describe SCs, associated with composite excitations built from paired electrons. This observation prompts the question of whether the BdG quantum geometry can directly reflect the internal pair structure, as hinted at in previous studies \cite{Tanaka2023,Daido2024,Simon2025}. Concretely, how does SC pairing, along with the normal state, contribute to the BdG quantum geometry? Are there in fact conditions under which we can identify a pairing quantum geometry, separable from the normal-state quantum geometry?

Beyond this conceptual motivation, studying BdG quantum geometry also has practical importance for superconductivity. On the one hand, topological superconductivity is commonly studied using symmetry-based classification schemes \cite{Schnyder2008,Kitaev2009,Ryu2010,Chiu2016}. However, these methods only identify the type of topological phase, e.g. $\mathbb{Z}$ or $\mathbb{Z}_{2}$, but not the value of its invariant, including it being zero or finite, to which only the BdG quantum geometry gives us access. Only explicit formulas for the BdG quantum geometry would provide a clear route toward designing topological SCs, through interplay between normal-state electronic structure and pairing, thereby establishing geometry-based design principles. On the other hand, unconventional superconductivity is notoriously difficult to characterize experimentally \cite{Goll2006, Stewart2017, Mackenzie2017, Aoki2022, Mandal2023}. The BdG quantum geometry, including the quantum metric, has already been shown to offer distinctive signatures \cite{Tanaka2023,Banerjee2026}, with more likely surfacing, just as quantum geometry has been recently shown to appear in a plethora of physical responses in the normal state \cite{Hasan2010,Ando2013,Xiao2006,Oka2009,Xiao2010,Liu2025,Jiang2025,Gao2025,Komissarov2024,Sala2026}.

A prerequisite for any of these developments is a general and systematic understanding of the structure of the BdG quantum geometry and, particularly, of its relation to the normal-state and pairing degrees of freedom.

In this work, we study the BdG quantum geometry of a wide range of SCs. Importantly, we establish sufficient conditions under which the BdG quantum geometry separates into a sum of the normal-state quantum geometry and an additional quantum geometry stemming purely from the SC pairing, hereafter called the {\it pairing quantum geometry}. These conditions involve three well-known concepts for multiorbital SCs. The first condition is SC fitness, which quantifies the compatibility of the normal state and the pairing matrix. SC fitness is intrinsically linked to the absence of interband pairing \cite{Ramires2018} and influences a range of properties, including odd-frequency pairing only appearing at finite fitness \cite{Triola2020}. Importantly, fit superconductors maximize the critical temperature \cite{Ramires2016,Ramires2018} and are therefore expected to be common amongst real SCs.
The second condition is orbital uniformity of the pairing, which occurs when the pairing matrix is proportional to the identity matrix in orbital space. This condition is routinely used in the context of understanding the (normal-state) quantum metric contributions to flat-band superconductivity \cite{Peotta2015,Tovmasyan2016}. Finally, the third condition is simply the absence of spin-flip terms in the normal state. All these three conditions are widely used, assumed satisfied by many SCs, and in many cases even produce the simplest SC pairing structures.

We show that when these three conditions are met, the BdG eigenstate is a product state of the normal state eigenstate and a pairing eigenstate. From this we establish that the BdG QGT separates as a sum of the normal state and pairing QGTs, thus displaying a simple, separable composite structure. Importantly, we provide explicit analytical formulas for the pairing quantum geometry for all spin-singlet and -triplet pairing possibilities, including the nonunitary case. This substantially extends earlier results, which have only focused on the limited case of time-reversal symmetric (TRS) spin-singlet pairing \cite{Porlles2023}. 
Our results, and especially the pairing quantum geometry, allow for a deeper understanding of the origins of topological superconductivity. For example, we show that in the spin-singlet case, a finite pairing Berry curvature arises from the coupling of the momentum variations of phase of the order parameter, with either that of the band dispersion or the amplitude of the order parameter. For spin-triplet pairings, especially nonunitary, the possibilities for nontrivial topology are even greater.

Finally, in order to explore the necessity of the conditions for a separable BdG quantum geometry, we also consider the case of unfit and non-uniform pairing in generic two-orbital spin-singlet SCs, assuming TRS. By analytically deriving the BdG quantum geometry, we show that the quantum geometry of the normal state and pairing are then generally intertwined in multiple different terms, whose origins we highlight. This result not only gives confidence in the necessity of SC fitness and orbital uniformity for a separable BdG quantum geometry, but, importantly, also highlights even vaster possibilities of nontrivial topology and finite quantum metric in SC systems. Our results can be directly applied to known SCs to determine their quantum geometry and, importantly, they provide a route for designing key quantum geometric properties, directly applicable to SC hybrid structures, which provide separate control of the normal state and the SC pairing.

The rest of this work is organized as follows. In Sec.~\ref{Sec:Notations} we recapitulate the notations of the multiorbital BdG framework. In Secs.~\ref{Sec:UnifitSinglet} and \ref{Sec:UnifitTriplet} we analytically derive the BdG quantum geometry and show its separability for all spin-singlet and -triplet pairing possibilities, respectively, under the three conditions discussed above. In Sec.~\ref{Sec:Non-Unifit} we compute the BdG quantum geometry of a general two-orbital spin-singlet SC without these conditions. In Sec.~\ref{Sec:Conclusion} we conclude our work.

\section{Setup: Multiorbital Superconductors}
\label{Sec:Notations}
In this work, we consider generic multiorbital SCs, with multiple active (i.e.~low-energy) orbitals in each unit cell, which can arise from sublattice of other internal degrees of freedom.
We let $N$ be the number of orbitals. Including also the spin degree of freedom, the normal state Hamiltonian is denoted as $H_{0}(\bk)$ and the SC pairing matrix as $\hat{\Delta}(\bk)$, both represented as $2N$ by $2N$ complex matrices. The BdG Hamiltonian $H(\bk)$ is then given by the following $4N$-dimensional matrix,
\begin{equation}
    H(\bk)=\begin{pmatrix}
        H_{0}(\bk)&\hat{\Delta}(\bk)\\
        \hat{\Delta}(\bk)^{\dagger}&-H_{0}(-\bk)^{*}
    \end{pmatrix}.
\end{equation}
Note that even in the single-orbital case, the BdG Hamiltonian has multiple degrees of freedom, namely spin and particle-hole, such that it can host quantum geometry \cite{Torma2023}. Throughout this work, we neglect the presence of spin-flip terms in the normal state, such that $H_{0}(\bk)=h_{\uparrow}(\bk)\oplus h_{\downarrow}(\bk)$. We note that this additional assumption is actually not needed for spin-triplet pairing, as it is one of the conditions for them to be fit, as discussed in Sec.~\ref{Sec:UnifitTriplet}. For spin-triplet pairing, only two conditions are then sufficient to ensure separability of the BdG quantum geometry. 

Considering both spin-singlet and -triplet SC pairing, we can generally write
\begin{equation}
    \hat{\Delta}(\bk)=\big[\hat{\psi}(\bk)\sigma_{0}+\hat{\bd}(\bk)\cdot\bm{\sigma}\big]i\sigma_{y}.
\end{equation}
where $\sigma_0$ and $\sigma_i$ are the identity and Pauli matrices, respectively, in spin space. Both $\hat{\psi}$ and $\hat{\bd}=(\hat{d}_{x},\hat{d}_{y},\hat{d}_{z})$ are $N$ by $N$ matrices, representing the spin-singlet and -triplet SC order parameters, respectively. The multiplication between a matrix in the orbital subspace and a Pauli matrix is to be taken as a tensor product. In this work, we consider pairing with a fixed spin-parity, either spin-singlet or spin-triplet. We leave the case of mixed spin-parity pairing for further study.

As we establish in this work, the BdG quantum geometry becomes separable into a sum of its normal state and SC pairing contributions when three conditions are met, two of which influences the SC pairing. The first such condition is that the SC is fit. 
The SC fitness $F$ is defined as \cite{Ramires2018}
\begin{equation}
    F(\bk)=H_{0}(\bk)\hat{\Delta}(\bk)-\hat{\Delta}(\bk)H_{0}(-\bk)^{*},
    \label{eq:Fitness-def}
\end{equation}
such that a SC is said to be fit if and only if (iff) $F(\bk)=0$.  A fit SC has been shown to maximize the SC critical temperature and is therefore considered to be a natural condition in many SCs \cite{Ramires2016,Ramires2018}. An unfit SC has $F(\bk)\neq 0$ for at least some momentum $\bk$, which we refer to simply as finite SC fitness (parameter).
The second condition is orbital uniformity for the pairing. The pairing is said to be orbital uniform iff $\hat{\psi}(\bk)=\psi(\bk)\mathbbm{1}$ and $\hat{\bd}(\bk)=\bd(\bk)\mathbbm{1}$. This means that the pairing is purely intraorbital and identical in each orbital. While a wide diversity of models have symmetries enforcing this condition, we note that it is not universal \cite{Tovmasyan2016}, still it is often used to derive at least analytical results \cite{Peotta2015,Iskin2023}.

Finally, we recall that the QGT associated with a quantum state $\ket{\Psi}$ is defined as
\begin{equation}
    Q^{\Psi}_{\mu\nu}=\braket{\pd{\mu}\Psi}{\pd{\nu}\Psi}-\braket{\pd{\mu}\Psi}{\Psi}\braket{\Psi}{\pd{\nu}\Psi}=g^{\Psi}_{\mu\nu}-\frac{i}{2}\mc{B}^{\Psi}_{\mu\nu},
    \label{eq:def-QGT}
\end{equation}
where $g^{\Psi}_{\mu\nu}$ is the quantum metric and $\mc{B}^{\Psi}_{\mu\nu}$ the Berry curvature \cite{Provost1980,Berry1989}. This assumes nondegeneracy of the energy eigenvalues. A generic degenerate set of eigenvectors geometry is instead described by the non-Abelian QGT:
\begin{equation}
    Q_{\mu\nu}^{\Psi} = \pd{\mu}\Psi^\dagger\pd{\nu}\Psi - \pd{\mu}\Psi^\dagger\Psi\Psi^\dagger\pd{\nu}\Psi 
    \label{eq:NAQGT_def}
\end{equation}
where $\Psi = (\ket{\Psi_1},\cdots,\ket{\Psi_n})$ are the horizontally stacked degenerated eigenvectors. The non-Abelian quantum metric and non-Abelian Berry curvature are defined by $g_{\mu\nu}^{\Psi} = (Q_{\mu\nu}^{\Psi}+Q_{\mu\nu}^{\Psi\dagger})/2$ and $B_{\mu\nu}^{\Psi} = i(Q_{\mu\nu}^{\Psi}-Q_{\mu\nu}^{\Psi\dagger})$, respectively \cite{Ma2010}.
In the following, we make extensive use of expressions for the QGT of unnormalized quantum states, derived in Appendix \ref{App:Shortcuts-Unnormalized}. 

\section{Fit and orbital uniform spin-singlet superconductivity}
\label{Sec:UnifitSinglet}
We start by deriving the BdG quantum geometry of spin-singlet SCs, assuming only three conditions. The normal state is diagonal in spin, the SC is fit and has orbital uniform pairing, such that $\hat{\Delta}(\bk)=i\hat{\psi}(\bk)\sigma_y$ with $\hat{\psi}(\bk)=\psi(\bk)\mathbbm{1}$, and decomposing $\psi(\bk) = |\psi(\bk)|\exp(i\varphi(\bk))$. 
It is easy to show that the fitness of the SC then holds at any momentum $\bk$ iff $\psi(\bk)$ vanishes, i.e.~ superconductivity is absent, or $h_{\uparrow}(\bk)=h_{\downarrow}(-\bk)^{*}$, i.e.~the normal state has time-reversal symmetry (TRS). We thus implement normal-state TRS at all momenta, such that the normal state Hamiltonian reads
\begin{equation}
    H_{0}(\bk)=\begin{pmatrix}
        h(\bk)&0\\
        0&h(-\bk)^{*}
    \end{pmatrix},
    \label{eq:NSmatrix-TRS}
\end{equation}
with $h=h_{\uparrow}$. Upon a rearrangement of the Nambu basis, the BdG Hamiltonian reads
\begin{equation}
    H(\bk)=\begin{pmatrix}
        h(\bk)&\psi(\bk)\mathbbm{1}&0&0\\
        \psi(\bk)^{*}\mathbbm{1}&-h(\bk)&0&0\\
        0&0&h(-\bk)^{*}&-\psi(\bk)\mathbbm{1}\\
        0&0&-\psi(\bk)^{*}\mathbbm{1}&-h(-\bk)^{*}
    \end{pmatrix}.
\end{equation}
The two diagonal blocks can be shown to be conjugated by a particle-hole transformation, such that the eigenstates of the upper block fully determine those of the lower block. Consequently, if $\ket{\phi}$ and $\ket{\psi}$ are eigenstates of the upper and lower block, respectively, their associated quantum geometric tensors can be shown to observe $Q^{\psi}_{\mu\nu}(\bk)=Q^{\phi}_{\mu\nu}(-\bk)^{*}$. We can therefore, without loss of information, focus on the upper block which, by abuse of notation, we denote as $H$. 

The eigenvalues are the usual spin-singlet-BdG eigenvalues $E_{n\eta}=\eta\sqrt{\epsilon_{n}^{2}+\abs{\psi}^{2}}$, with $\epsilon_{n}$ the normal state dispersion, $\eta=\pm1$, and where the $\bk$-dependence is implicit in the following. As for the eigenvectors, denoted as $\ket{\Psi_{n\eta}}$, it is easy to show that (see Supplementary Material (SM) \cite{SM})
\begin{equation}
    \ket{\Psi_{n\eta}}=\begin{pmatrix}
        u_{n\eta}\\
        v_{n\eta}
    \end{pmatrix}\otimes\ket{\text{u}_{n}}.
    \label{eq:singlettensor}
\end{equation}
Here $\ket{\text{u}_{n}}$ are the normal state eigenstates, i.e.~$h\ket{\text{u}_{n}} = \epsilon_n \ket{\text{u}_{n}}$ and the SC coherence factors are given by
\begin{equation}
    u_{n\eta}=\frac{1}{\sqrt{2}}\sqrt{1+\frac{\epsilon_{n}}{E_{n\eta}}},\hspace{2mm}v_{n\eta}=\eta\frac{e^{-i\varphi}}{\sqrt{2}}\sqrt{1-\frac{\epsilon_{n}}{E_{n\eta}}}.
\end{equation}
With the tensor product structure of the BdG eigenstate in Eq.~\eqref{eq:singlettensor}, we find by using Appendix \ref{App:Shortcuts-Product} that the associated BdG QGT exactly separates into the sum of the normal-state quantum geometry, associated with $\ket{\text{u}_{n}}$, and an additional pairing quantum geometry associated with the SC coherence factors,
\begin{equation}
Q^{n\eta}_{\mu\nu}=Q^{\text{u}_{n}}_{\mu\nu}+Q^{\text{p}n\eta}_{\mu\nu}.
\label{eq:QGT-separ-singlet}
\end{equation}
This reveals the separable and composite nature of the quantum geometry for SCs, with two simple and distinct contributions. On the one hand, the normal-state quantum geometry is associated with the electrons composing the Cooper pairs. On the other hand, the pairing quantum geometry emerges from the attractive interaction between these components.
A direct and important consequence of this separability is that the ensuing topology and quantum metric of the SC come from either or both the normal state and the pairing. As such, even relatively simple spin-singlet SCs can acquire finite quantum geometry through two different means. Note here that considering the BdG Hamiltonian in the normal-state band basis would neglect the contribution of the normal state to the quantum geometry of the SC. 

Since normal-state quantum geometry is already well studied, we here instead focus on the pairing quantum geometric tensor, keeping the normal-state band index $n$ implicit in the following. By applying Eq.~\eqref{eq:def-QGT} to the vector of coherence factors in Eq.~\eqref{eq:singlettensor}, we arrive at the pairing quantum geometry (see SM \cite{SM})
\begin{equation}
    Q_{\mu\nu}^{\text{p}\eta} = \dfrac{1}{4E_\eta^4}T^{\eta}_{\mu} T^{\eta*}_{\nu},
    \label{eq:QGT-singlet}
\end{equation}
with $T_\mu^\eta = |\psi|\partial_\mu\epsilon-\epsilon\pd{\mu}|\psi|-iE_\eta|\psi|\partial_\mu\varphi$. A generic property of pairing quantum geometry is that it diverges when the order parameter vanishes, as the BdG bands then become degenerate \cite{Porlles2023}. Below we study separately the pairing quantum metric and Berry curvature.

\subsection{Pairing quantum metric}
We first consider the pairing quantum metric $g^{\text{p}\eta}_{\mu\nu}=\Re Q^{\text{p}\eta}_{\mu\nu}$, given by
\begin{multline}
    g^{\text{p}\eta}_{\mu\nu}=\frac{\abs{\psi}^{2}}{4E_{\eta}^{4}}\pd{\mu}\epsilon\pd{\nu}\epsilon+\frac{\epsilon^{2}}{4E_{\eta}^{4}}\pd{\mu}\abs{\psi}\pd{\nu}\abs{\psi}\\-\frac{\epsilon\abs{\psi}}{4E_{\eta}^{4}}\pd{(\mu}\epsilon\pd{\nu)}\abs{\psi}+\frac{\abs{\psi}^{2}}{4E_{\eta}^{2}}\pd{\mu}\varphi\pd{\nu}\varphi,
    \label{eq:Singletpairingmetric}
\end{multline}
where we have introduced the symmetric tensor notation
\begin{equation}
    A_{(\mu}B_{\nu)}=A_{\mu}B_{\nu}+A_{\nu}B_{\mu}.
\end{equation}
The first three terms were previously derived in Ref.~\cite{Porlles2023}, which focused on only single-orbital TRS spin-singlet SC. The first term is present in all SCs, provided the normal state band is dispersive, and is reminiscent of the conventional contribution to the superfluid weight \cite{Peotta2015}. The second term is purely related to the variations (in momentum space) of the order parameter amplitude $|\psi|$, while the third term couples it with the band dispersion $\epsilon$. Thus, the second term appears whenever the SC pair amplitude, or equivalently here the SC gap, is momentum dependent, and the third term additionally appears if the normal state band is not flat. The last term is only present for a momentum dependent SC phase $\varphi$ and thus only appears when the spin-singlet SC breaks TRS. This last term may therefore be used to probe TRS breaking in spin-singlet SCs \cite{Ghosh2020}. Physically, since the quantum metric is linked to the delocalization of the associated quantum state as it is the spread of Wannier functions \cite{Marzari1997,Resta2011,Komissarov2024,Marsal2024,Simon2025}, our results establish how the normal-state, SC gap, and phase momentum modulations combined contribute to the spread of the BdG quasiparticle wave functions.

\subsection{Pairing Berry curvature}
Next, focusing on the pairing Berry curvature $\mc{B}^{\text{p}\eta}_{\mu\nu}=-2\Im Q^{\text{p}\eta}_{\mu\nu}$, we obtain (see SM \cite{SM})
\begin{equation}
\mc{B}^{\text{p}\eta}_{\mu\nu}=\frac{\abs{\psi}}{2E_{\eta}^{3}}\big(\epsilon\pd{[\mu}\abs{\psi}\pd{\nu]}\varphi-\abs{\psi}\pd{[\mu}\epsilon\pd{\nu]}\varphi\big),
    \label{eq:SingletpairingBerry}
\end{equation}
with the antisymmetric tensor notation
\begin{equation}
    A_{[\mu}B_{\nu]}=A_{\mu}B_{\nu}-A_{\nu}B_{\mu}.
\end{equation}
Here we directly see that the pairing Berry curvature is only present when the spin-singlet order parameter breaks TRS and then appears when the momentum variation of the phase $\varphi$ couples either to the band dispersion $\epsilon$ or the SC gap $\abs{\psi}$. As a consequence, these are two additional sources of topological superconductivity for orbital uniform and fit spin-singlet SCs, in addition to any topology generated by the normal state. In the case of a flat band, only the latter comes into play.

We note here that we can extend the picture of Ref.~\cite{Porlles2023} in which the quantum metric was interpreted as the texture of the vector field $(v_{n\eta}, -u_{n\eta})$. We find that this perspective is in fact related to the pseudospin interpretation of BCS theory proposed by Anderson \cite{Anderson1958}. In this interpretation, the Hamiltonian is expressed as $H=\mathbf{h}\cdot\bm\tau$ where $\bm\tau$ is the vector of Pauli matrices representing pseudospin operators, and $\hat{\mathbf{h}}=\mathbf{h}/E_\eta=(\Re(\psi),-\Im(\psi),\epsilon)/E_\eta$ defines the pseudospin. Then, the hybridization between the electron and hole bands, due to superconductivity, corresponds to the formation of a pseudospin domain wall at the Fermi surface, separating an up and a down phase. A rotation of the pseudospin about the $x$- or $y$-axis corresponds to a variation of either $\epsilon$ or $\psi$, while a rotation about the $z$-axis corresponds to a variation of $\varphi$ only. As a consequence, we find that the full quantum geometry, i.e.~both the quantum metric and Berry curvature, is associated with the texture of the pseudospin, via the usual expression of the quantum geometric tensor for a two-level system \cite{Graf2021}, here expressed through the Bloch vector $\hat{\mathbf{h}}$.

\section{Fit and orbital uniform spin-triplet superconductivity}
\label{Sec:UnifitTriplet}
We next consider spin-triplet SCs. As we see below, the additional degrees of freedom provided by the spin-triplet state give rise to a richer structure of the QGT. An orbital uniform spin-triplet state is described by $\hat{\Delta}(\bk) = i\hat{\bd}(\bk)\cdot\bm{\sigma}\sigma_y$ with $\hat{\bd}(\bk) = \bd(\bk)\mathbbm{1}$. The order parameter $\bd$ observes TRS iff its components are real. There are two types of spin-triplet order parameter depending on the form of $\Delta(\bk)\Delta^\dagger(\bk) = |\bd(\bk)|^2\sigma_0+\mathbf{q}(\bk)\cdot\bm\sigma$: a nonunitary case where $\mathbf{q}(\bk)\neq0$ and a unitary case where $\mathbf{q}(\bk)=0$ and thus $\Delta(\bk)\Delta^\dagger(\bk)\propto\sigma_0$. Specifically, unitary pairing is associated with spin degeneracy, as in the spin-singlet case, while nonunitary pairing lifts the spin degeneracy and thus breaks TRS.

We again consider a fit and orbital uniform SC.
Requiring fitness, i.e. $F=0$ in Eq.~\eqref{eq:Fitness-def}, of spin-triplet orbital uniform pairing is equivalent to three conditions on the normal state Hamiltonian $H_0$. First, the absence of spin-flip terms in $H_0$, which we already assumed. Second, it requires TRS, such that $h_\uparrow(\bk) = h_\downarrow(-\bk)^*$, and, finally, spin degeneracy with $h_\uparrow(\bk) = h_\downarrow(\bk)$. The normal-state Hamiltonian thus reads $H_0(\bk) = h(\bk)\sigma_0$ and the BdG Hamiltonian reduces to 
\begin{equation}
    H(\bk) = \begin{pmatrix}
        h(\bk)\sigma_0 & i\bd(\bk)\cdot\bm\sigma\sigma_y\mathbbm{1} \\ 
        -i\sigma_y\bd(\bk)^*\cdot\bm\sigma\mathbbm{1} & -h(\bk)\sigma_0
    \end{pmatrix}.
    \label{eq:HBdG-triplet}
\end{equation}
We here note that for simplicity we consider normal state Hamiltonians that are fit for all spin-triplet order parameters $\bd$. Considering a specific $\bd$ vector may lead to weaker conditions, for example, if the pairing is purely opposite spin or equal spin. Making the $\mathbf{k}$ dependence implicit in the following, the eigenvalues of $H$ in Eq.~\eqref{eq:HBdG-triplet} are $\eta E_{n\sigma} = \eta\sqrt{\epsilon_n^2+|\bd|^2+\sigma q}$, where $q = |\mathbf{q}|$ and $\eta,\sigma=\pm 1$ differentiate energies in the particle-hole and spin sectors, respectively. The corresponding eigenvectors decompose into the tensor product of the normal state eigenstates $\ket{\text{u}_n}$ and now with a vector in Nambu-spin space  $\ket{\chi_{n\eta\sigma}}$ (see SM \cite{SM})
\begin{equation}
    \ket{\Psi_{n\eta\sigma}} = \mathcal{N}_{n\sigma} \ket{\chi_{n\eta\sigma}} \otimes \ket{\text{u}_n}, \quad \ket{\chi_{n\eta\sigma}}=\begin{pmatrix}
        u_{n\eta\sigma}  \\
        v_{n\eta\sigma}
    \end{pmatrix}
    \ket{\phi_{n\eta\sigma}}.
    \label{eq:Triplet-Eigst-separ}
\end{equation}
Here, $\ket{\chi_{n\eta\sigma}}$ corresponds to the eigenstates in the single orbital case \cite{Sigrist1991}, while $u_{n\eta\sigma}$, $v_{n\eta\sigma}$ are two-dimensional matrices given by
\begin{equation}
\begin{split}
    u_{n+\sigma}=v_{n-\sigma}&=(E_{n\sigma}+\epsilon_n)\sigma_0, \\  
    u_{n-\sigma}=v_{n+\sigma}^\dagger&=i\bd\cdot\bm\sigma\sigma_y,
    \label{eq:u_v_triplet}
\end{split}
\end{equation}
and $\mathcal{N}_{n\sigma} = [2E_{n\sigma}(E_{n\sigma}+\epsilon_n)]^{-1/2}$ is the normalization factor. Finally,  $\ket{\phi_{n\eta\sigma}}$ is a normalized two-dimensional vector that we specify below.

In the unitary case, spin degeneracy  $E_{n+}=E_{n-}=E_{n}$ implies spin degeneracy of $u_{n\eta\sigma}$, $v_{n\eta\sigma}$ and $\mc{N}_{n\sigma}$, and freedom in the determination of the eigenvectors. We here arbitrarily choose $u_{n+\sigma}=v_{n-\sigma}\propto\sigma_0$ in Eq.~\eqref{eq:u_v_triplet}. In addition, we arbitrarily choose $\ket{\phi_{n\eta\sigma}} = (\sigma+1,\sigma-1)^\intercal/2$ to be an eigenvector of $\sigma_z$.
In the nonunitary case, the spin degeneracy is lifted, which removes the arbitrariness in the determination of the eigenstates. The state $\ket{\phi_{n\eta\sigma}}$ is now determined by the eigenvalue equations
\begin{equation}
\hat{\mathbf{q}}\cdot\bm\sigma\ket{\phi_{n+\sigma}} = \sigma\ket{\phi_{n+\sigma}}, \quad\hat{\mathbf{q}}\cdot\bm\sigma^*\ket{\phi_{n-\sigma}} = \sigma\ket{\phi_{n-\sigma}},
\label{eq:eigpb-q}
\end{equation}
where $\hat{\mathbf{q}}=\mathbf{q}/q$.

Just as in the spin-singlet case in Sec.~\ref{Sec:UnifitSinglet}, the separability of the eigenstates in Eq.~\eqref{eq:Triplet-Eigst-separ} ensures the separability of the QGT into a normal state and a pairing contribution
\begin{equation}
    Q_{\mu\nu}^{n\eta(\sigma)} = Q_{\mu\nu}^{\text{u}_n} + Q_{\mu\nu}^{\text{p}n\eta(\sigma)}.
\end{equation}
The spin label $\sigma$ is here relevant only in the nondegenerate nonunitary case. Below, we analytically derive the pairing quantum geometry for both unitary and nonunitary pairings, for a given band index $n$, which we make implicit in the following. Moreover, the symmetry relation between the particle and hole eigenstates $\ket{\Psi_{+\sigma}(\mathbf{k})} = i\tau_y\ket{\Psi_{-\sigma}(\mathbf{k})}^*$, implies that $Q_{\mu\nu}^{+(\sigma)}(\mathbf{k})=Q_{\mu\nu}^{-(\sigma)}(\mathbf{k})^*$, which allows us to consider only the particle sector QGT in the nonunitary case. In the unitary case, the particle and hole sectors can be dealt with simultaneously, as seen below.

\subsection{Unitary spin-triplet pairing}
We start by investigating unitary spin-triplet SCs and their pairing contribution to the QGT. Due to spin degeneracy, we need to consider the non-Abelian QGT $Q_{\mu\nu}^{\text{p}\eta}$ in Eq.~\eqref{eq:NAQGT_def} with eigenstates $\chi_{\eta}=(\ket{\chi_{\eta+}},\ket{\chi_{\eta-}})$. Furthermore, in the unitary case, $\mathbf{q}=0$, which implies $\bd \propto \bd^*$, i.e. $\bd$ is proportional to a real vector. We can thus redefine $\bd \rightarrow \bd e^{i\varphi}$, where $\bd\in\mathbb{R}^3$ and $\varphi\in\mathbb{R}$ is a global phase. The QGT then reads (see SM \cite{SM})
\begin{equation}
    Q_{\mu\nu}^{\text{p}\eta} = \dfrac{1}{4E^4}T^{\eta}_\mu T^{\eta\dagger}_\nu,
    \label{eq:QGT_uni}
\end{equation}
with $T^\eta_\mu = d\partial_\mu\epsilon-\epsilon\hat{\bd}\cdot\partial_\mu\bd+i\eta E(\hat{\bd}\times\partial_\mu\bd\cdot\bm\sigma-d\partial_\mu\varphi)$, where $\hat{\bd}=\bd/d$ and $d=|\bd|$. We omit the identity matrix $\sigma_0$ here and in the following for scalars that are proportional to it. The QGT has large similarities with the spin-singlet case in Eq.~\eqref{eq:QGT-singlet} if we replace $d$ with $|\psi|$, except for the notable additional term $\hat{\bd}\times\pd{\mu,\nu}\bd\cdot\bm\sigma$. This term generates the non-Abelian structure by introducing off-diagonal components via $\bm\sigma$. 

\subsubsection{Pairing quantum metric} Symmetrizing the QGT tensor in Eq.~\eqref{eq:QGT_uni}, we obtain the non-Abelian pairing quantum metric
\begin{multline}
    g_{\mu\nu}^{\text{p}\eta} = \dfrac{1}{4E^4}(d^2\pmu\epsilon \pnu\epsilon+ \epsilon^2\pmu d\pnu d - \epsilon d \pd{(\mu}\epsilon \pd{\nu)}d) \\ 
    + \dfrac{1}{4E^2}(d^2\pmu\varphi\pnu\varphi +d^2\pmu\hat{\bd}\cdot\pnu\hat{\bd} -\partial_{(\mu}\varphi\bd\times\partial_{\nu)}\bd\cdot\bm\sigma).
    \label{eq:Unitary-metric}
\end{multline}
The terms proportional to the identity can be interpreted as the Abelian sector, while the $\bm\sigma$ terms constitute the non-Abelian sector, indicating that the geometrical structure is dependent on the spin projection. The first three terms involve the order parameter only via its amplitude $d$ and are analogous to the spin-singlet case. Likewise, the fourth term, dependent on the phase of the $\bd$-vector is analogous to the spin-singlet case.
However, the three-dimensional structure of the spin-triplet vector provides two additional geometrical terms, involving dot or cross products of $\bd$. The fifth term still belongs to the Abelian sector and is the pairing quantum metric associated with the two-level system $\bd\cdot\bm{\sigma}$ \cite{Graf2021}, and is thus only present for spin-triplet pairing due to its additional degrees of freedom. Finally, the sixth and last term is the only non-Abelian contribution stemming from the spin degeneracy of spin-triplet unitary pairing. It couples momentum modulations of $\varphi$ with the spin-space twisting of $\bd$. The cross product $\bd\times\pnu\bd$ is finite for any generic nonuniform momentum space texture of $\hat{\bd}$. Notably, the sixth term is only non-zero for a momentum dependent $\varphi$ and, as such, is only there when the pairing breaks TRS.

\subsubsection{Pairing Berry curvature}
Instead, antisymmetrizing the QGT in Eq.~\eqref{eq:QGT_uni} yields the non-Abelian Berry curvature
\begin{multline}
    \mc{B}_{\mu\nu}^{\text{p}\eta} = \dfrac{\eta d}{2E^3}\left(\epsilon\partial_{[\mu}d-d\partial_{[\mu}\epsilon\right)\partial_{\nu]}\varphi + \dfrac{\eta}{2E^3}\partial_{[\mu}\epsilon\bd\times\partial_{\nu]}\bd\cdot\bm\sigma \\
    - \dfrac{\eta\epsilon}{2E^3}\pmu\bd\times\pnu\bd\cdot\bm\sigma - \dfrac{\eta}{2E^3(E+\epsilon)}\left(\bd\cdot\pmu\bd\times\pnu\bd\right)\bd\cdot\bm\sigma. 
    \label{eq:Unitary-curvature}
\end{multline}
The Berry curvature is generated by the momentum space texture of the order parameter. The first term belongs to the Abelian sector and is the direct analog of the spin-singlet pairing Berry curvature in Eq.~\eqref{eq:SingletpairingBerry}. It originates from the interplay between a global phase twist of $\varphi$ in momentum space and momentum variations of the pairing amplitude $d$. In contrast, the remaining terms are of non-Abelian origin and are entirely controlled by the momentum-space texture of the $\bd$ vector. As a consequence, they remain finite even in the absence of any global phase variation. 
The second term couples the normal-state to the texture of the order parameter through the normal-state dispersion $\partial_{\mu}\epsilon$ and therefore vanishes in the flat-band limit. The last two terms depend solely on the geometry of the $\bd$ texture and persist even for dispersionless bands. In particular, the third term is generated by any finite twisting of $\bd$ in momentum space, whereas the fourth term is proportional to the Berry curvature of the effective two-level system $\bd\cdot\bm{\sigma}$ and therefore probes the three-dimensional winding of the order parameter \cite{Graf2021}. A key consequence of Eq.~\eqref{eq:Unitary-curvature} is that, unlike in the spin-singlet case, a finite pairing Berry curvature does not require a momentum-dependent phase $\varphi$, but a nontrivial texture of the $\bd$ vector alone is sufficient to generate it. Therefore, pairing-induced Berry curvature, and thereby non-trivial topology set by the integrated Berry curvature, arise much more generically, even in unitary spin-triplet SCs. Notably, this is entirely independent of any Berry curvature contribution that may stem from the normal state, as this is topology generated by SC pairing alone.

\subsection{Nonunitary spin-triplet pairing}
We next turn to nonunitary spin-triplet pairing and extract the pairing QGT, which in this case is Abelian due to the spin splitting induced by the finite $\bf{q}$. It is also sufficient to consider only the eigenstates of the particle sector $\ket{\Phi_{+\sigma}}$. The derivation (see SM \cite{SM}) provides
\begin{equation}
    Q_{\mu\nu}^{\text{p}+\sigma} = \mathcal{N}_{\sigma}^2A_{\mu\nu} - \mathcal{N}_{\sigma}^4B_{\mu}B_{\nu}^* + Q_{\mu\nu}^\phi,
    \label{eq:QGT-nonunitary}
\end{equation}
where 
\begin{align}
        A_{\mu\nu} =& \Lambda_\mu\Lambda_\nu + \mathcal{D}_\mu^\sigma\bd\cdot\pd{\nu}\bd^* +\tfrac{\sigma}{2}\left[i\mathcal{D}_\mu^\sigma\hat{\mathbf{q}}\cdot\bm\pi_\nu+ \text{H.c.}\right], \nonumber \\
        B_{\mu} =& (E_{\sigma}+\epsilon)\Lambda_{\mu}+\bd^*\cdot\mathcal{D}_\mu^\sigma\bd, \nonumber \\
        Q_{\mu\nu}^\phi =& \tfrac{1}{4}(\pmu\hat{\mathbf{q}}\cdot\pnu\hat{\mathbf{q}}+i\sigma\hat{\mathbf{q}}\cdot\pmu\hat{\mathbf{q}}\times\pnu\hat{\mathbf{q}}),
        \label{eq:QGT_nonuni_terms}
\end{align}
with $E_{\sigma}\Lambda_{\mu}=(E_{\sigma}+\epsilon)\pd{\mu}\epsilon+\Re(\bd^*\cdot\mathcal{D}_\mu^\sigma\bd)$, where $\mathcal{D}_\mu^\sigma=\pd{\mu}+i\sigma\hat{\mathbf{q}}\times\pd{\mu}$, and $\bm\pi_\mu=\Re(\bd\times\pd{\mu}\bd^*)$.  The last term $Q_{\mu\nu}^\phi$ is the standard QGT of the two-band system for which the nonunitary spin eigenstates $\ket{\phi_{+\sigma}}$, given in Eq.~\eqref{eq:eigpb-q}, form a basis \cite{Graf2021}. This is a purely nonunitarity driven quantum geometry. At the same time, $\Lambda_\mu\in\mathbb{R}$, so $\Lambda_\mu\Lambda_\nu$, appearing in both $A_{\mu\nu}$ and $B_{\mu}B_\nu^*$, is real and thereby associated only with the quantum metric. Other terms in $A_{\mu\nu}$ and $B_{\mu}B_{\nu}^*$ are complex valued and more difficult to interpret. These terms all involve the operator $\mathcal{D}_\mu^\sigma$ that takes the derivative of a vector and projects its transverse component into the helicity eigenstate defined by $\hat{\mathbf{q}}$, i.e. the eigenvectors $\mathbf{e}_\sigma$ satisfying $i\hat{\mathbf{q}}\times\mathbf{e}_\sigma=\sigma\mathbf{e}_\sigma$. 

\subsubsection{Pairing quantum metric} The quantum metric is found as the real part of the QGT in Eq.~\eqref{eq:QGT-nonunitary}
as 
\begin{widetext}
\begin{multline}
g^{\text{p}+\sigma}_{\mu\nu}=\frac{\abs{\bd}^{2}+\sigma q}{4E_{\sigma}^{4}}\pd{\mu}\epsilon\pd{\nu}\epsilon-\frac{(E_{\sigma}+\epsilon)^{2}+E_{\sigma}^{2}}{16E_{\sigma}^{4}(E_{\sigma}+\epsilon)^{2}}\pd{\mu}\big(\abs{\bd}^{2}+\sigma q\big)\pd{\nu}\big(\abs{\bd}^{2}+\sigma q\big)-\frac{\epsilon}{8E_{\sigma}^{4}}\pd{(\mu}\epsilon\pd{\nu)}(\abs{\bd}^{2}+\sigma q)\\
    +\frac{\mc{N}_{\sigma}^2}{2}\bigg[\pd{(\mu}\bd\cdot\pd{\nu)}\bd^{*}+i\sigma\pd{(\mu}\bd\times\pd{\nu)}\bd^{*}\cdot\hat{\mathbf{q}}+\hat{\mathbf{q}}\times\pd{(\mu}\hat{\mathbf{q}}\cdot\bm{\pi}_{\nu)}-2\mc{N}_\sigma^2(\alpha_{\mu}+\bm{\pi}_{\mu}\cdot\hat{\mathbf{q}})(\alpha_{\nu}+\bm{\pi}_{\nu}\cdot\hat{\mathbf{q}})\bigg]+\frac{1}{4}\pd{\mu}\hat{\mathbf{q}}\cdot\pd{\nu}\hat{\mathbf{q}},
    \label{eq:Nonunitary-metric}
\end{multline}
\end{widetext}
where $\alpha_\mu=\Im(\bd\cdot\pd{\mu}\bd^*)$.
The first three terms are reminiscent of the first three terms in the spin-singlet case in Eq.~\eqref{eq:Singletpairingmetric}. The first term comes from the normal state dispersion, thus being zero for flat bands. The second term measures the contribution from the pairing amplitude variations in momentum space, while the third term mixes the two previous contributions. A notable feature of the general TRS breaking nonunitary state is the presence of the spin-splitting of the quasiparticle energies $E_{n\sigma}$, given by $\sigma q$. A variation of $q$ is also associated with variations of the amplitudes and relative phases of the components of the order parameter $\bd$. Thus, $q$ becomes an integral part of the quantum metric and appears (indirectly) in every term in the quantum metric. The remaining terms also encode both deformations of the amplitudes and relative phases of the order parameter $\bd$. In particular, the last term is the quantum metric associated with the two-level system $\hat{\mathbf{q}}\cdot\bm\sigma$ and measures the variation of the spin-polarization axis $\hat{\mathbf{q}}$.

\subsubsection{Pairing Berry curvature} The imaginary part of the QGT in Eq.~\eqref{eq:QGT-nonunitary} yields the Berry curvature
\begin{widetext}
\begin{multline}
    \mc{B}^{p+\sigma}_{\mu\nu}=\frac{1}{2E_{\sigma}^{3}}\left[\pd{[\mu}\epsilon+\frac{2E_{\sigma}+\epsilon}{2(E_{\sigma}+\epsilon)^{2}}\pd{[\mu}\big(\abs{\bd}^{2}+\sigma q\big) \right]\left[\alpha_{\nu]}+\sigma\bm{\pi}_{\nu]}\cdot\hat{\mathbf{q}}\right]\\
    +\mc{N}_{\sigma}^2\left(i\pd{[\mu}\bd\cdot\pd{\nu]}\bd^{*}-\sigma\pd{[\mu}\bd\times\pd{\nu]}\bd^{*}\cdot\hat{\mathbf{q}}+\sigma\bm{\pi}_{[\mu}\cdot\pd{\nu]}\hat{\mathbf{q}}\right)-\frac{\sigma}{2}\hat{\mathbf{q}}\cdot\pd{\mu}\hat{\mathbf{q}}\times\pd{\nu}\hat{\mathbf{q}}.
    \label{eq:Nonunitary-curvature}
\end{multline}
\end{widetext}  
As in the unitary case, the Berry curvature is generated by the momentum-space texture of the complex order parameter. In the nonunitary case this texture involves not only the SC phase but also the relative phases and amplitudes of the components of $\bd$, as well as the polarization structure encoded in $\mathbf{q}$. In fact, all terms in $\mc{B}_{\mu\nu}^{\text{p}+\sigma}$ contain contributions of both amplitude and phase variations. We further identify the last term as the standard Berry curvature associated with the two-level system $\hat{\mathbf{q}}\cdot\bm{\sigma}$ \cite{Graf2021}. Overall, the pairing Berry curvature  demonstrates the multitude of possibilities of arriving at non-trivial topology stemming entirely from the nonunitary SC order parameter.

To gain some intuition of Eqs.~\eqref{eq:QGT-nonunitary}, \eqref{eq:Nonunitary-metric} and \eqref{eq:Nonunitary-curvature}, we consider the simple example of an order parameter of the form $\mathbf{d}=\Delta(1,ik_z,0)$, which has only real-valued equal spin pairing. For this choice, $\mathbf{q}=(0,0,2\Delta^2k_z)$ so $\hat{\mathbf{q}}=\hat{\mathbf{z}}$ and the state is further characterized by a fixed relative phase of $\pi/2$ between the $x$- and $y$-components, while the relative amplitude varies with $k_z$. Consequently, the spin polarization vector $\mathbf{q}$ has a fixed direction but momentum-dependent amplitude. Since the order parameter depends only on $k_z$, the only nonvanishing component of the pairing QGT is $Q_{zz}^{\text{p}+\sigma}=g_{zz}^{\text{p}+\sigma}$, i.e.~only a finite quantum metric. Moreover, $\partial_z\hat{\mathbf{q}}=0$, implying $Q_{\mu\nu}^\phi=\mc{D}_\mu^\sigma\hat{\mathbf{q}}=0$, so that the last term in $A_{zz}$ in Eq.~\eqref{eq:QGT_nonuni_terms} vanishes. The first three terms of the quantum metric in Eq.~\eqref{eq:Nonunitary-metric} remain finite with $\partial_z(|\bd|^2+\sigma q)=2\Delta^2(k_z+\sigma)$. Among the remaining terms, only $\pd{z}\bd\cdot\pd{z}\bd^{*}=\Delta^2$ and $(\bm\pi_z\cdot\hat{\mathbf{q}})^2=\Delta^4$ are nonzero. This simple example thus illustrates that a momentum-dependent magnitude of the nonunitary vector alone is sufficient to generate a nontrivial quantum metric. By contrast, remaining geometric contributions, and in particular a finite Berry curvature, require a more intricate momentum-space texture of the $\bd$-vector, but those cases come from a multitude of relatively simple possibilities: a dependence on multiple momentum components, a momentum-dependent relative or global phase, or a momentum-dependent spin-polarization axis.

Finally, we note that the limit from the nonunitary to the unitary case is ill-defined, due to the energy degeneracy of the unitary states. This is seen in practice in the appearance of the indeterminate limit of $\hat{\mathbf{q}}=\mathbf{q}/q$ when $q\rightarrow 0$. However, we can obtain the unitary limit by considering instead the non-Abelian QGT for the nonunitary eigenstates spanning the spin subspaces, and only then taking the limit of zero $q$.

\section{Unfit and orbital non-uniform superconductivity}
\label{Sec:Non-Unifit}
Having derived analytical expressions for the contribution from SC pairing to the QGT for SC fit and orbital uniform SCs, we finally turn to relaxing these two conditions, while still assuming the absence of spin-flip terms in the normal state.
To be able to still find an analytically tractable system, we consider the simplest BdG Hamiltonian that is neither orbital uniform nor fit. As we establish below, the normal state and pairing contributions are then generically no longer separable and the BdG quantum geometry is thus much richer, as the two contributions intertwine. The benefit of considering a simple model is also that we then effectively establish the necessity of fitness and orbital uniform pairing for the separation of the normal state and pairing contributions to the QGT.

Specifically, we consider a SC with two-orbital spin-singlet pairing and time-reversal symmetry. Denoting the identity and Pauli matrices in the orbital subspace as $(\alpha_{0},\bm{\alpha})$, the spin-singlet pairing matrix reads
\begin{equation} \hat{\psi}=\psi_{0}\alpha_{0}+\bm{\psi}\cdot\bm{\alpha}.
\end{equation}
This decomposition allows us to separate the orbital uniform part in $\psi_{0}$ from the non-uniform part in $\bm{\psi}$.
TRS further constraints the pairing matrix to be Hermitian, such that $\psi_{0}$ and $\bm{\psi}$ are both real quantities.  As for the normal state, TRS constrains $H_{0}$ to be given by Eq.~(\ref{eq:NSmatrix-TRS}), i.e., $H_{0}(\bk)=\hat{h}(\bk)\oplus\hat{h}(-\bk)^{*}$. 
We also decompose $\hat{h}(\bk)=h_{0}(\bk)\alpha_{0}+\mathbf{h}(\bk)\cdot\bm{\alpha}$.
The associated SC fitness $F(\bk)$ can be expressed as
\begin{equation}
    F(\bk)=2i\begin{pmatrix}
        0&\mathbf{f}(\bk)\cdot\bm{\alpha}\\
        \mathbf{f}(-\bk)\cdot\bm{\alpha}^{*}&0
    \end{pmatrix},
\end{equation}
with the vector $\bf{f}(\bk)=\bf{h}(\bk)\times\bm{\psi}(\bk)$, which we call the fitness vector. The SC is then unfit iff it has a finite fitness vector $\bf{f}(\bk)$. In the orbital uniform case, the fitness vector vanishes and thus orbital uniform pairing here automatically falls within the framework of Sec.~\ref{Sec:UnifitSinglet}.
In the rest of this section, we make the momentum dependencies implicit.

By the same argument as in Sec. \ref{Sec:UnifitSinglet}, the BdG Hamiltonian decomposes into a direct sum of two blocks, where the quantum geometry of one block directly determines that of the other. Using the same abuse of notation as in Sec. \ref{Sec:UnifitSinglet}, we focus on the upper block and denote the resulting BdG Hamiltonian 
\begin{equation}
    H=\begin{pmatrix}
        \hat{h}&\hat{\psi}\\
        \hat{\psi}&-\hat{h}
    \end{pmatrix}.
    \label{eq:H_unfit_nonuniform}
\end{equation}
 A crucial consequence of TRS is that the BdG Hamiltonian in Eq.~\eqref{eq:H_unfit_nonuniform}, being inherently particle-hole symmetric, also exhibits chiral symmetry. By applying the unitary matrix
\begin{equation}
    U=\frac{1}{\sqrt{2}}\begin{pmatrix}
        i&i\\
        1&-1
    \end{pmatrix}\otimes\alpha_{0},
\end{equation}
the BdG Hamiltonian becomes off-diagonal,
\begin{equation}
    U^{\dagger}HU=\begin{pmatrix}
        0&\hat{Q}^{\dagger}\\
        \hat{Q}&0
    \end{pmatrix},\hspace{2mm}\hat{Q}=\hat{h}-i\hat{\psi}.
\end{equation}
The BdG eigenvalues are then directly determined from those of $\hat{Q}\hat{Q}^{\dagger}$. Computing the characteristic polynomial yields the four BdG eigenvalues,
\begin{equation}
    E_{\eta\eta'}=\eta\sqrt{h_{0}^{2}+h^{2}+\psi_{0}^{2}+\psi^{2}+2\eta'v},
\end{equation}
with $\eta,\eta'=\pm1$, $h=\abs{\mf{h}}$, $\psi=\abs{\bm{\psi}}$, and $v=\abs{h_{0}\mathbf{h}+\psi_{0}\bm{\psi}+\mathbf{f}}$. By manipulating the eigenproblem in the chiral basis, we arrive at the following BdG eigenstates \cite{SM},
\begin{equation}
    \ket{\Psi_{\eta\eta'}}=\frac{1}{\sqrt{2}E_{\eta\eta'}}\begin{pmatrix}
        E_{\eta\eta'}\ket{\phi_{\eta'}}\\
        Q\ket{\phi_{\eta'}}
    \end{pmatrix},
    \label{eq:eigenstate_unfit_nonuniform}
\end{equation}
where the quantum state $\ket{\phi_{\eta'}}$ obeys the eigenproblem $Q^{\dagger}Q\ket{\phi_{\eta'}}=E_{\eta\eta'}^{2}\ket{\phi_{\eta'}}$, which is equivalent to
\begin{equation}
    \mathbf{v}_{+}\cdot\bm{\alpha}\ket{\phi_{\eta'}}=\eta'v\ket{\phi_{\eta'}}.
\end{equation}
with $\mf{v}_{+}=\mf{w}+
\mf{f}$, where $\mf{w}=h_{0}\mf{h}+\psi_{0}\bm{\psi}$.

Setting $\ket{\psi_{1}}=E_{\eta\eta'}\ket{\phi_{\eta'}}$ and $\ket{\psi_{2}}=Q\ket{\phi_{\eta'}}$, Eq.~\eqref{eq:eigenstate_unfit_nonuniform} appears as the concatenation of two unnormalized quantum states of equal norms. Using this structure of the BdG eigenstates in Eq.~\eqref{eq:eigenstate_unfit_nonuniform}, we show in Appendix \ref{App:Shortcuts-Concatenated} that the BdG QGT can be decomposed as
\begin{multline}
    Q^{\eta'}_{\mu\nu}=\frac{1}{2}Q^{\psi_{1}}_{\mu\nu}+\frac{1}{2}Q^{\psi_{2}}_{\mu\nu}\\+\frac{1}{4n^{4}}(A^{\psi_{1}}_{\mu}-A^{\psi_{2}}_{\mu})^{*}(A^{\psi_{1}}_{\nu}-A^{\psi_{2}}_{\nu}),
    \label{eq:QGT-concat}
\end{multline}
with $n^{2}=\braket{\psi_{i}}=E_{\eta\eta'}^{2}$. We here introduce the shorthand notation $A^{\psi_{i}}_{\mu}=i\braket{\psi_{i}}{\pmu\psi_{i}}$, which can be viewed as a Berry pseudoconnection, as the states $\ket{\psi_{i}}$ are not normalized. Moreover, we find that the BdG QGT does not depend on the band index $\eta$, and we thus denote the BdG quantum geometry only using the band index $\eta'$. The BdG QGT is then purely determined by, first, the average of the QGTs of $\ket{\psi_{1}}$ and $\ket{\psi_{2}}$ individually, and, second, by a term resulting from the mismatch of their Berry pseudoconnections.  As shown in the SM \cite{SM}, the mismatch term is real, implying that it contributes only to the BdG quantum metric. This also means that the Berry curvature is given by the averaged Berry curvature of $\ket{\psi_{1}}$ and $\ket{\psi_{2}}$. Below, we separately consider the BdG quantum metric and Berry curvature.

\subsection{BdG quantum metric}
The BdG quantum metric is given by the real part of the QGT $Q^{\eta'}_{\mu\nu}$ in Eq.~\eqref{eq:QGT-concat}. The full derivation, see SM  \cite{SM}, yields
\begin{equation}
    g^{\eta'}_{\mu\nu}=\frac{1}{4}\pd{\mu}\hat{\mf{b}}\cdot\pd{\nu}\hat{\mf{b}}+\frac{1}{4b^{2}E_{\eta\eta'}^{4}}T_{\mu}^{\eta'}T_{\nu}^{\eta'},
    \label{eq:Metric-Nonunifit}
\end{equation}
with $\mathbf{b}=(\mathbf{w},\mathbf{f})$, $\hat{\mathbf{b}}=\mathbf{b}/b$, and
\begin{multline}
    T_{\mu}^{\eta'}=(v\psi_{0}+\eta'\mf{w}\cdot\bm{\psi})\pd{\mu}h_{0}+(v\bm{\psi}+\eta'\psi_{0}\mf{w}+\eta'\mf{f}\times\mf{h} )\cdot\pd{\mu}\mf{h}\\-(vh_{0}+\eta'\mf{w}\cdot\mf{h})\pd{\mu}\psi_{0}\\-(v\mf{h}+\eta'h_{0}\mf{w}-\eta'\mf{f}\times\bm{\psi})\cdot\pd{\mu}\bm{\psi}.
\end{multline}
The first term in the quantum metric in Eq.~\eqref{eq:Metric-Nonunifit} stems from the average of the quantum metrics of $\ket{\psi_{1}}$ and $\ket{\psi_{2}}$, while the second term comes from the mismatch of their Berry pseudoconnections. An important takeaway of Eq.~(\ref{eq:Metric-Nonunifit}) is that although we reach a closed form expression in terms of $\mathbf{w}$ and $\mathbf{f}$, there exist no clear structure with regards to the normal state and pairing matrices. Instead, the normal state and pairing contributions to the quantum metric are inextricably mixed, such that the separability of the BdG quantum geometry, proven for orbital uniform and fit pairings in Sections \ref{Sec:UnifitSinglet} and \ref{Sec:UnifitTriplet}, no longer holds. This non-separability of the BdG quantum geometry, even in such a simple situation as a two-orbital spin-singlet SC without orbital uniformity and SC fitness gives confidence in the necessity of these two conditions for a separable composite BdG quantum geometry, outside possible fine-tuned normal states and pairing matrices, where terms may accidentally still cancel each other. This also directly establishes that a finite BdG quantum geometry is easily generated in cases of finite fitness and orbital non-uniform pairing, as it can stem not only from individual contributions from the normal state or the pairing, but also from intertwined structures. From the point of view of quantum geometry, the individual components of the Cooper pair, the electrons, described by $H_0$, and their attractive interaction, described by $\hat{\Delta}$, thus become entangled such that the resulting Cooper pair has a non-separable composite character.

Moreover, the finite fitness vector $\bf{f}$ also has two identifiable effects in Eq.~(\ref{eq:Metric-Nonunifit}). It weakens the amplitude of the quantum metric through the denominator, while also producing additional terms. To better see this second effect, let us set $h_{0}=\psi_{0}=0$, such that $\mf{w}=0$. The BdG quantum metric then becomes
\begin{equation}
    g^{\eta'}_{\mu\nu}=\frac{1}{4}\pd{\mu}\hat{\mathbf{f}}\cdot\pd{\nu}\hat{\mathbf{f}}+\frac{1}{4f^{2}E_{\eta\eta'}^{4}}T_{\mu}^{\eta'}T_{\nu}^{\eta'},
    \label{eq:Metric-SecV-fitnessdriven}
\end{equation}
where
\begin{equation}
    T_{\mu}^{\eta'}=\frac{1}{2}f\pd{\mu}(h^{2}-\psi^{2})+\eta'\mf{f}\cdot(\pd{\mu}\mf{h}\times\bm{\psi}-\mf{h}\times\pd{\mu}\bm{\psi}),
\end{equation}
with $f=|\mf{f}|$ and $\hat{\mf{f}}=\mf{f}/f$.
The first term in Eq.~\eqref{eq:Metric-SecV-fitnessdriven} is the quantum metric associated to the two-level Hamiltonian $\mf{f}\cdot\bm{\alpha}$, and thus constitutes a purely finite fitness-driven quantum metric in a two-orbital superconductor. The second term stems from the difference of momentum variations between the vectors $\mf{h}$ and $\bm{\psi}$, both in magnitude and direction in the orbital space. It is also controlled by the finite fitness vector $\mf{f}$. Having a finite SC fitness is thus generically sufficient to drive a non-zero BdG quantum metric. 
Conversely, if we set the fitness vector $\bf{f}$ to zero, which also includes the cases ${\bm h} = 0$ or ${\bm \psi} =0$, but still keep $\mf{w}$ finite, the quantum metric reads
\begin{equation}
    g^{\eta'}_{\mu\nu}=\frac{1}{4}\pd{\mu}\hat{\mf{w}}\cdot\pd{\nu}\hat{\mf{w}}+\frac{1}{4w^{2}E_{\eta\eta'}^{4}}T_{\mu}^{\eta'}T_{\nu}^{\eta'},
    \label{eq:metric-SecV-fit}
\end{equation}
with $w=|\mf{w}|$ and $\hat{\mf{w}}=\mf{w}/w$. A finite quantum metric is then generated by the vector $\bf{w}$. In particular, in the orbital uniform case where $\bm{\psi}=0$, the first term of Eq.~\eqref{eq:metric-SecV-fit} becomes the normal state quantum metric associated with $\mf{h}\cdot\bm{\alpha}$ \cite{Graf2021}, while the second term reproduces the pairing quantum metric of Eq.~\eqref{eq:Singletpairingmetric} in the TRS case. This establishes how results in this Section directly connect to those in Section~\ref{Sec:UnifitSinglet} for fit and orbital uniform pairing.
Finally, setting the pairing matrix $\hat{\psi}$ to zero makes $T_{\mu}^{\eta'}$ vanish and we recover the normal-state quantum metric.

\subsection{BdG Berry curvature}
The BdG Berry curvature is given by the imaginary part of the QGT $Q^{\eta'}_{\mu\nu}$ in Eq.~\eqref{eq:QGT-concat}, where only the first term contributes, as shown in the SM \cite{SM}. Consequently, we have $\mc{B}^{\eta'}_{\mu\nu}=(\mc{B}_{\mu\nu}^{\psi_{1}}+\mc{B}_{\mu\nu}^{\psi_{2}})/2$. In the supplementary material we show that the Berry curvatures $\mc{B}^{\psi_{1}}_{\mu\nu}$ and $\mc{B}^{\psi_{2}}_{\mu\nu}$ are the Berry curvatures associated to the two-level Hamiltonians $\mf{v}_{+}\cdot\bm{\alpha}$ and $\mf{v}_{-}\cdot\bm{\alpha}$, respectively, with $\mf{v}_{\pm}=\mf{w}\pm\mf{f}$. Using the usual two-band formula for the two terms leads to \cite{Graf2021}
\begin{equation}
    \mc{B}^{\eta'}_{\mu\nu}=-\frac{\eta'}{2v^{3}}\big(\mathbf{v}_{+}\cdot\pd{\mu}\mathbf{v}_{+}\times\pd{\nu}\mathbf{v}_{+}+\mathbf{v}_{-}\cdot\pd{\mu}\mathbf{v}_{-}\times\pd{\nu}\mathbf{v}_{-}\big).
\end{equation}
In terms of $\bf{w}$ and $\bf{f}$, this becomes \cite{SM}
\begin{widetext}
    \begin{equation}
        \mc{B}^{\eta'}_{\mu\nu}=-\frac{\eta'}{2(w^2+f^2)^{3/2}}\big(\mathbf{w}\cdot\pd{\mu}\mathbf{w}\times\pd{\nu}\mathbf{w}+\mathbf{w}\cdot\pd{\mu}\mathbf{f}\times\pd{\nu}\mathbf{f}+\mathbf{f}\cdot\pd{\mu}\mathbf{w}\times\pd{\nu}\mathbf{f}+\mathbf{f}\cdot\pd{\mu}\mathbf{f}\times\pd{\nu}\mathbf{w}\big).
        \label{eq:BerryNonUnifit}
    \end{equation}
\end{widetext}
The fitness vector $\bf{f}$ has two clearly identifiable effects. First, it weakens the magnitude of the Berry curvature as it appears in the denominator of the prefactor. Second, it is the source of additional terms in the Berry curvature. However, setting $\mathbf{w}=0$ results in a vanishing BdG Berry curvature. This is achieved for example in the case of a uniform pairing with $h_{0}=0$. In contrast to the BdG quantum metric, a finite fitness vector is thus not sufficient to produce a finite BdG Berry curvature. 

We get further insight into Eq.~(\ref{eq:BerryNonUnifit}) by considering the fit case, $\mathbf{f}=0$, as this implies that $\mathbf{h}$ and $\bm{\psi}$ are locked in the orbital subspace, such that we can write $\bm{\psi}=\lambda\mathbf{h}$, for a scalar $\lambda$. We then have
\begin{equation}
    \mc{B}^{\eta'}_{\mu\nu}=-\sgn(h_{0}+\lambda\psi_{0})\frac{\eta'}{2h^{3}}\mathbf{h}\cdot\pd{\mu}\mathbf{h}\times\pd{\nu}\mathbf{h},
    \label{eq:BerryNonUnifit-fit}
\end{equation}
which is just the normal-state Berry curvature up to a sign factor. Thus, orbital non-uniform pairing in the fit case only influences the Berry curvature via a possible sign flip, depending on its strength and sign with respect to $h_0$ and $\psi_0$. In particular, restricting Eq.~(\ref{eq:BerryNonUnifit-fit}) to the uniform orbital pairing where $\lambda=0$, we simply recover the normal-state Berry curvature up to the sign of $h_{0}$. This sign factor stems from the fact that if $h_{0}$ changes sign, then the normal state bands are swapped and the Berry curvature also changes sign. The normal state band index is then given by $\eta'\sgn(h_{0})$. This is expected, as TRS for spin-singlet SCs together with it both being fit and having uniform pairing, makes the pairing Berry curvature vanish.

\section{Conclusions}
\label{Sec:Conclusion}
In this work we establish explicit formulas for the quantum geometry of SCs, dubbed the BdG quantum geometry. 
We show that it generally develops a composite structure with contributions from both the normal state quantum geometry and a quantum geometry stemming from the SC order parameter, dubbed the pairing quantum geometry. In particular, we provide sufficient conditions under which the BdG quantum geometry exactly separates into the sum of the normal state and pairing quantum geometries. One of these conditions is orbital uniformity of the pairing, meaning the pairing is purely intraorbital with the same order parameter in each orbital. Another condition is SC fitness, which links the normal state and pairing matrices in a modified commutation relation. Finally, we also neglect the presence of spin-flip terms in the normal state Hamiltonian. These conditions are assumed to be fulfilled in many SCs.

We further provide explicit analytical formulas for the pairing quantum geometry for all spin-singlet and -triplet SCs, including nonunitary states, under the conditions discussed above. For spin-singlet SCs, the pairing quantum geometry is given by Eqs.~(\ref{eq:Singletpairingmetric}) and (\ref{eq:SingletpairingBerry}). Notably, the pairing quantum metric is non-trivial even in the simplest momentum independent $s$-wave case, provided the normal state band is dispersive. If instead the normal state band is flat, pairing quantum metric originates, separately, from momentum dependence of either the amplitude or the phase of the order parameter. The pairing Berry curvature, and hence topology, on the other hand, is non-trivial only when the spin-singlet order parameter breaks time-reversal symmetry as it requires momentum dependence of the SC phase.

For spin-triplet SCs, we find a richer structure due to the vector character of the order parameter $\bd$. For unitary pairing, the spin degeneracy of the quasiparticle energies leads to a non-Abelian pairing quantum geometry, expressed in Eqs.~\eqref{eq:Unitary-metric} and \eqref{eq:Unitary-curvature}. For both the pairing quantum metric and Berry curvature, we find terms directly analogous to the spin-singlet case and with additional terms originating from the vector character of the $\bd$-vector. In the quantum metric, there are two such additional terms, one being the quantum metric associated with the two-level system $\bd\cdot\bm{\sigma}$, being Abelian. and another one involving the phase of the $\bd$-vector, being non-Abelian. In contrast, the Berry curvature has a more pronounced non-Abelian character, including one term involving the Berry curvature associated with 
the two level system $\bd\cdot\bm{\sigma}$. An important observation is that, while in the spin-singlet case a momentum dependent phase of the order parameter is needed to produce a finite Berry curvature, the vector character of the momentum-dependent $\bd$-vector is enough in the spin-triplet case. We also treat the nonunitary case, where the spin-polarization of the pairing lifts the spin degeneracy and leads to an Abelian pairing quantum geometry established in Eqs.~\eqref{eq:Nonunitary-metric} and \eqref{eq:Nonunitary-curvature}. The expressions we find in the nonunitary case are more complicated, which makes their interpretation more involved. In particular, the QGT associated with the nonunitary vector $\mathbf{q}$ appears, thus establishing the presence of a purely non-unitarity driven quantum geometry. Overall these results provide concrete evidence that quantum geometry and thus topology are very common in spin-triplet SCs.

Lastly, having proven the sufficiency of orbital uniformity and SC fitness to have separability of the BdG quantum geometry, with a separate pairing quantum geometry, we explore a simple case beyond these conditions to both study their necessity and illustrate how finite SC fitness and orbital nonuniformity easily contributes to the QGT. Specifically, we analytically compute the BdG quantum geometry of a two-orbital spin-singlet SC with time-reversal symmetry. Our results show that the normal state and the pairing contributions generally intertwine in the QGT. Thus, neglecting fine-tuned cases where parameters may possibly be chosen such that terms accidentally cancel, these results give confidence in the necessity of orbital uniformity and SC fitness for a pure pairing QGT. We further establish that having a nonzero SC fitness alone generates a finite BdG quantum metric.

Our results provide a strong foundation for further theoretical and experimental works on topological and unconventional superconductivity. In particular, they highlight the multitude possibilities to produce quantum geometry in SCs, in particular, stemming just from the SC pairing, even in such simple SCs as those that are both fit and with uniform orbital pairing.

\section*{Acknowledgments}
We thank A.~Bhattacharya and Q.~Marsal for useful discussions. We acknowledge financial support from the Swedish Research
Council (Vetenskapsrådet) Grant No.~2022-03963 and the European Union through the European Research Council (ERC) under the European Union’s Horizon 2020 research and innovation programme (ERC-2022-CoG, Grant agreement No.~101087096). Views and opinions expressed are, however, those of the authors only and do not necessarily reflect those of the European Union or the European Research Council Executive Agency. Neither the European Union nor the granting authority can be held responsible for them. F.S. also acknowledges financial support from the Technical University of Munich, under the TUM Global Postdoc Fellowship.

\bibliography{Biblio1}

\appendix
\section{Shortcuts to computing quantum geometry}
\label{App:Shortcuts}
In this Appendix, we derive the QGT for specific classes of quantum states, which we use throughout the paper. In this Appendix, $\ket{\phi}$ and $\ket{\psi}$ denote two generic quantum states. The QGT associated with $\ket{\psi}$ is defined as \cite{Berry1989},
\begin{equation}
    Q^{\psi}_{\mu\nu}=\braket{\pd{\mu}\psi}{\pd{\nu}\psi}-\braket{\pd{\mu}\psi}{\psi}\braket{\psi}{\pd{\nu}\psi}.
    \label{eq:defQGT}
\end{equation}
Introducing the Berry connection $\mc{A}^{\psi}_{\mu}=i\braket{\psi}{\pd{\mu}\psi}$, we also have
\begin{equation}
    Q^{\psi}_{\mu\nu}=\braket{\pd{\mu}\psi}{\pd{\nu}\psi}-\mc{A}^{\psi}_{\mu}\mc{A}^{\psi}_{\nu}.
    \label{eq:QGT-Berrycon}
\end{equation}

The QGT of the set of eigenvectors $\ket{\psi_i}$ of a degenerate space is given by the non-Abelian generalization
\begin{equation}
    Q_{\mu\nu}^{ab} = \braket{\pmu\psi_a}{\pnu\psi_b} - \sum_c \braket{\pmu\psi_a}{\psi_c}\braket{\psi_c}{\pnu\psi_b}.
\end{equation}
It is compactly written in matrix form introducing the vector of horizontally stacked eigenvectors $\psi=(\ket{\psi_1}\cdots\ket{\psi_2})$ as
\begin{equation}
    Q_{\mu\nu} = \pd{\mu}\psi^\dagger\pd{\nu}\psi-\pd{\mu}\psi^\dagger\psi\psi^\dagger\pd{\nu}\psi.
\end{equation}

\subsection{QGT of product quantum states}
\label{App:Shortcuts-Product}
We show that the QGT of a tensor product of quantum states is separable. This result is used Sections \ref{Sec:UnifitSinglet} and \ref{Sec:UnifitTriplet}. Consider the product state of $\ket{\phi}$ and $\ket{\psi}$,
\begin{equation}
    \ket{\Psi}=\ket{\phi}\otimes\ket{\psi},
\end{equation}
where we omit the tensor product symbol in the following. Its derivative trivially reads
\begin{equation}
    \ket{\pd{\mu}\Psi}=\ket{\pd{\mu}\phi}\ket{\psi}+\ket{\phi}\ket{\pd{\mu}\psi}.
\end{equation}
As such,
\begin{equation}
    \braket{\pd{\mu}\Psi}{\pd{\nu}\Psi}=\braket{\pd{\mu}\phi}{\pd{\nu}\phi}+\braket{\pd{\mu}\psi}{\pd{\nu}\psi}+\mc{A}^{\psi}_{\mu}\mc{A}^{\phi}_{\nu}+\mc{A}^{\psi}_{\nu}\mc{A}^{\phi}_{\mu}.
\end{equation}
As for the Berry connection, we simply have
\begin{equation}
    \mc{A}^{\Psi}_{\mu}=\mc{A}^{\phi}_{\mu}+\mc{A}^{\psi}_{\mu}.
\end{equation}
Plugging everything in the QGT then yields
\begin{align}
    Q^{\Psi}_{\mu\nu}&=\braket{\pd{\mu}\psi}{\pd{\nu}\psi}+\braket{\pd{\mu}\phi}{\pd{\nu}\phi}+\mc{A}^{\psi}_{\mu}\mc{A}^{\phi}_{\nu}+\mc{A}^{\psi}_{\nu}\mc{A}^{\phi}_{\mu}\nonumber\\
    &-(\mc{A}^{\phi}_{\mu}+\mc{A}^{\psi}_{\mu})(\mc{A}^{\phi}_{\nu}+\mc{A}^{\psi}_{\nu}).
\end{align}
The cross-terms then cancel out and we obtain the separability of the QGT of product states,
\begin{subequations}
\begin{align}
    Q^{\Psi}_{\mu\nu}&=\braket{\pd{\mu}\psi}{\pd{\nu}\psi}+\braket{\pd{\mu}\phi}{\pd{\nu}\phi}-\mc{A}^{\psi}_{\mu}\mc{A}^{\psi}_{\nu}-\mc{A}^{\phi}_{\mu}\mc{A}^{\phi}_{\nu}\\
    &=Q^{\psi}_{\mu\nu}+Q^{\phi}_{\mu\nu}.
\end{align}
\end{subequations}

For a non-Abelian QGT, we consider the vector $\Psi = \ket{\phi}\otimes\psi$, where $\psi=(\ket{\psi_1}\cdots\ket{\psi_n})$. Then replacing $\ket{\psi}\rightarrow\psi$, and defining $\mathcal{A}_\mu^\psi=i\psi^\dagger\pd{\mu}\psi$, the derivation is identical and yields the same separability of the QGT. 

\subsection{QGT of unnormalized quantum states}
\label{App:Shortcuts-Unnormalized}
Let us consider an unnormalized quantum state $\Ket{\tilde{\psi}}$. The resulting QGT depends only on the derivative of $\Ket{\tilde{\psi}}$ and not on those of the normalization factor, which greatly simplifies the calculations of the pairing quantum geometries in Sections \ref{Sec:UnifitSinglet} and \ref{Sec:UnifitTriplet}. Let $\mc{N}=\Braket{\tilde{\psi}}{\tilde{\psi}}^{-1/2}$ be its normalization factor, and $\ket{\psi}=\mc{N}\Ket{\tilde{\psi}}$ its normalized state. Its derivative then reads
\begin{equation}
    \ket{\pd{\mu}\psi}=\pd{\mu}\mc{N}\Ket{\tilde{\psi}}+\mc{N}\Ket{\pd{\mu}\tilde{\psi}}.
\end{equation}
On the one hand, we obtain
\begin{align}
    \braket{\pd{\mu}\psi}{\pd{\nu}\psi}&=\mc{N}^{-2}\pd{\mu}\mc{N}\pd{\nu}\mc{N}+\mc{N}\pd{\mu}\mc{N}\Braket{\tilde{\psi}}{\pd{\nu}\tilde{\psi}}\nonumber\\
    &+\mc{N}\pd{\nu}\mc{N}\Braket{\pd{\mu}\tilde{\psi}}{\tilde{\psi}}+\mc{N}^{2}\Braket{\pd{\mu}\tilde{\psi}}{\pd{\nu}\tilde{\psi}}.
\end{align}
On the other hand, 
\begin{align}
    &\braket{\pd{\mu}\psi}{\psi}\braket{\psi}{\pd{\nu}\psi}=\mc{N}^{-2}\pd{\mu}\mc{N}\pd{\nu}\mc{N}+\mc{N}\pd{\mu}\mc{N}\Braket{\tilde{\psi}}{\pd{\nu}\tilde{\psi}}\nonumber\\
    &+\mc{N}\pd{\nu}\mc{N}\Braket{\pd{\mu}\tilde{\psi}}{\tilde{\psi}}+\mc{N}^{4}\Braket{\pd{\mu}\tilde{\psi}}{\tilde{\psi}}\Braket{\tilde{\psi}}{\pd{\nu}\tilde{\psi}}.
\end{align}
We see that when subtracting the two equations to obtain the QGT,  all the terms involving derivatives of $\mc{N}$ cancel out. Expliciting the latter, we obtain the QGT associated with an unnormalized quantum state,
\begin{equation}
    Q^{\psi}_{\mu\nu}=\frac{\Braket{\pd{\mu}\tilde{\psi}}{\pd{\nu}\tilde{\psi}}}{\Braket{\tilde{\psi}}{\tilde{\psi}}}-\frac{\Braket{\pd{\mu}\tilde{\psi}}{\tilde{\psi}}\Braket{\tilde{\psi}}{\pd{\nu}\tilde{\psi}}}{\Braket{\tilde{\psi}}{\tilde{\psi}}^{2}}.
    \label{eq:defQGT-notnorm}
\end{equation}
Mathematically, Eq. (\ref{eq:defQGT}) describes the quantum geometry of a $\text{U}(1)$-fiber bundle, while Eq. (\ref{eq:defQGT-notnorm}) encapsulates the quantum geometry associated with a $\mathbb{C}^{*}$-fiber bundle. 

For a non-Abelian QGT, we consider a set of unnormalized eigenvectors with equal norms $\Ket{\tilde{\psi}_i}$ spanning a degenerate space. The normalized eigenvectors thus read $\ket{\psi_i}=\mathcal{N}\Ket{\tilde{\psi}_i}$, with $\mathcal{N}=\Braket{\tilde{\psi}_i}{\tilde{\psi}_i}^{-1/2}$ for all $i$. Then the normalized vector $\psi=(\ket{\psi_1}\cdots\ket{\psi_n})=\mathcal{N}(\Ket{\tilde{\psi}_1}\cdots\Ket{\tilde{\psi}_n})=\mathcal{N}\tilde{\psi}$. Following the above derivation while replacing $\ket{\phi}\rightarrow\phi$ yields:
\begin{equation}
    Q_{\mu\nu}^\psi = \dfrac{\pd{\mu}\tilde{\psi}^\dagger\pd{\nu}\tilde{\psi}}{\tilde{\psi}^\dagger\tilde{\psi}}-\dfrac{\pd{\mu}\tilde{\psi}^\dagger\tilde{\psi}\tilde{\psi}^\dagger\pd{\nu}\tilde{\psi}}{(\tilde{\psi}^\dagger\tilde{\psi})^2}
\end{equation}
with $\tilde{\psi}^\dagger\tilde{\psi}=\Braket{\tilde{\psi}}{\tilde{\psi}}\mathbbm{1}_n$.

\subsection{QGT of concatenated quantum states}
\label{App:Shortcuts-Concatenated}
We derive the expression used in Section \ref{Sec:Non-Unifit} for the QGT of a quantum state $\ket{\Psi}$ defined as the concatenation of two unnormalized quantum states $\ket{\psi_{1}}$ and $\ket{\psi_{2}}$. For the application to Eq.~\eqref{eq:eigenstate_unfit_nonuniform}, it is sufficient to consider the two to have equal norms $n^{2}=\braket{\psi_{1}}=\braket{\psi_{2}}$. The normalized concatenated state then reads
\begin{equation}
    \ket{\Psi}=\frac{1}{\sqrt{2}n}\begin{pmatrix}
        \ket{\psi_{1}}\\
        \ket{\psi_{2}}
    \end{pmatrix}.
\end{equation}
Applying Eq.~\eqref{eq:defQGT-notnorm}, we have
\begin{multline}
    Q^{\Psi}_{\mu\nu}=\frac{\braket{\pd{\mu}\psi_{1}}{\pd{\nu}\psi_{1}}+\braket{\pd{\mu}\psi_{2}}{\pd{\nu}\psi_{2}}}{2n^2}-\frac{1}{4n^4}(A^{\psi_{1}}_{\mu}+A^{\psi_{2}}_{\mu})^{*}\\(A^{\psi_{1}}_{\nu}+A^{\psi_{2}}_{\nu}),
\end{multline}
where we introduced the shorthand notation $A^{\psi_{1}}_{\mu}=i\braket{\psi_{1}}{\pd{\mu}\psi_{1}}$. We also apply Eq.~\eqref{eq:defQGT-notnorm} to $\ket{\psi_{1}}$, which gives us
\begin{equation}
    \braket{\pd{\mu}\psi_{1}}{\pd{\nu}\psi_{1}}=n^{2}Q^{\psi_{1}}_{\mu\nu}+n^{-2}A^{\psi_{1}*}_{\mu}A^{\psi_{1}}_{\nu},
\end{equation}
and similarly for $\ket{\psi_{2}}$. The QGT of $\ket{\Psi}$ then reads
\begin{multline}
    Q^{\Psi}_{\mu\nu}=\frac{1}{2}Q^{\psi_{1}}_{\mu\nu}+\frac{1}{2}Q^{\psi_{2}}_{\mu\nu}+\frac{1}{2n^{4}}A^{\psi_{1}*}_{\mu}A^{\psi_{1}}_{\nu}+\frac{1}{2n^{4}}A^{\psi_{2}*}_{\mu}A^{\psi_{2}}_{\nu}\\-\frac{1}{4n^{2}}(A^{\psi_{1}}_{\mu}+A^{\psi_{2}}_{\mu})^{*}(A^{\psi_{1}}_{\nu}+A^{\psi_{2}}_{\nu}),
\end{multline}
which we factorize as follows,
\begin{equation}
    Q^{\Psi}_{\mu\nu}=\frac{1}{2}Q^{\psi_{1}}_{\mu\nu}+\frac{1}{2}Q^{\psi_{2}}_{\mu\nu}+\frac{1}{4n^{4}}(A^{\psi_{1}}_{\mu}-A^{\psi_{2}}_{\mu})^{*}(A^{\psi_{1}}_{\nu}-A^{\psi_{2}}_{\nu}).
\end{equation}

\end{document}